\documentstyle[12pt,epsf]{article}
\newcommand{\lsim}{\mathrel{\raisebox{-.6ex}{$\stackrel{\textstyle<}{\sim}$}}}
\newcommand{\gsim}{\mathrel{\raisebox{-.6ex}{$\stackrel{\textstyle>}{\sim}$}}}
\def\eps{\epsilon}
\def\ts{\textstyle}

\begin{document}

\font\fortssbx=cmssbx10 scaled \magstep2
\hbox to \hsize{
\hbox{\fortssbx University of Wisconsin - Madison}
      \hfill$\vcenter{
\hbox{\bf MADPH-97-1031}
\hbox{\bf AMES-HET-97-13}
       \hbox{December 1997}}$ }

\vspace{.5in}

\begin{center}
{\bf FOUR-WAY NEUTRINO OSCILLATIONS}
\\
\vskip 0.7cm
{V. Barger$^1$, T.J. Weiler$^{1,2}$, and K. Whisnant$^3$}
\\[.1cm]
$^1${\it Department of Physics, University of Wisconsin, Madison, WI
53706, USA}\\
$^2${\it Department of Physics and Astronomy, Vanderbilt University,
Nashville, TN 37235, USA}\\
$^3${\it Department of Physics and Astronomy, Iowa State University,
Ames, IA 50011, USA}\\
\end{center}

\smallskip

\begin{abstract}
We present a four-neutrino model with three active neutrinos and one
sterile neutrino which naturally has maximal $\nu_\mu\rightarrow\nu_\tau$
oscillations of atmospheric neutrinos and can explain the
solar neutrino and LSND results. The model predicts $\nu_e \rightarrow
\nu_\tau$ and $\nu_e \rightarrow \nu_\mu$ oscillations in long--baseline
experiments with $L/E \gg 1$~km/GeV with amplitudes that are determined
by the LSND oscillation amplitude and argument controlled by the atmospheric $\delta m^2$.

\end{abstract}

\thispagestyle{empty}
\newpage

There is growing experimental evidence that neutrinos
oscillate~\cite{review}. The long-standing solar neutrino
deficit~\cite{SSM,solar}, the atmospheric neutrino
anomaly~\cite{atmos,SuperK,oldatmos}, and the recent results from the
LSND experiment on neutrinos from $\mu^+$ and $\pi^+$ decay~\cite{LSND}
can all be understood in terms of oscillations between two neutrino
species. The challenge is to describe all oscillation phenomena within a
single model, since resonant oscillations for the sun, oscillations for
the atmosphere, and the LSND data each require a different neutrino
mass--squared difference $\delta m^2$ to properly describe all features
of the data~\cite{3nu}.  For example, if the atmospheric $\delta m^2$
scale is raised to the LSND scale, one forfeits the recently reported
zenith--angle dependence of the atmospheric neutrino flux~\cite{SuperK}.
Alternatively, if the solar $\delta m^2$ is raised to the atmospheric
$\delta m^2$ scale, one finds that: (i) the reduction in the solar
neutrino flux is energy--independent~\cite{flat}, and (ii) near--maximal
$\nu_e-\nu_\mu$ or $\nu_e-\nu_\tau$ mixing is necessary to describe the
observed solar $\nu_e$ suppression; but in the context of a $3\times3$
unitary matrix, such large mixing is inconsistent with the near maximal
$\nu_\mu-\nu_\tau$ mixing deduced from the atmospheric data. Then also
if $\delta m^2 \ge 10^{-3} {\rm~eV}^2$, large $\nu_e$ mixing with any
neutrino species is excluded by the recent CHOOZ reactor
data~\cite{CHOOZ}. Hence the large suppression of the solar flux can
only result from resonance--enhancement, which requires a very small
mass scale compared to those indicated by the atmospheric and LSND data.
Since three-neutrino models can have at most two independent
mass-squared differences $\delta m^2$, apparently not all the data can
be explained with just $\nu_e$, $\nu_\mu$, and $\nu_\tau$.

A viable solution is to postulate one or more additional species of
sterile light neutrino~\cite{VB80} without Standard Model gauge
interactions (to be consistent with LEP measurements of $Z
\rightarrow \nu\bar\nu$~\cite{Znunubar}) thereby introducing another
independent mass scale to the theory. The latter approach has been used
with some success in the literature~\cite{models,Mohap}. The constraints
of big-bang nucleosynthesis give the constraint
\begin{equation}
\delta m^2 A < 10^{-7} {\rm~eV}^2 \;,
\end{equation}
 on the mass-squared difference $\delta m^2$ and oscillation amplitude
$A=\sin^22\theta$ for oscillations between a sterile neutrino and an
active neutrino flavor~\cite{BBN}.

In this Letter, we examine a four-neutrino model (three active plus one
sterile\footnote
{ In principle, more than one sterile neutrino species can exist.
However, only the particular linear combination of sterile neutrinos
that mixes with $\nu_e$ is phenomenologically interesting.}
)
which naturally has maximal $\nu_\mu\rightarrow\nu_\tau$ oscillations of
atmospheric neutrinos and which can also explain the solar neutrino and
LSND results. We begin with a brief discussion of the three classes of
experiments and the neutrino mass and mixing parameters needed to
explain them. We then present a mass matrix whose eigenvalues
consist of a nearly degenerate neutrino pair at $\sim 1$~eV and a nearly
degenerate pair at low mass, as illustrated in Fig.~1. We show
how the existing data
almost uniquely fixes the model parameters and strictly determines what
new phenomenology the model predicts. We find that the new
observable signature for the model (in addition to the oscillations
already indicated by the data) is $\nu_e \leftrightarrow \nu_\mu$ and
$\nu_e \leftrightarrow \nu_\tau$ oscillations for $L/E \gg 1$~km/GeV.
We discuss the
possibility that such signals can be observed in long-baseline neutrino
experiments such as those using intense muon sources at
Fermilab~\cite{Geer} or KEK and detectors at SOUDAN, GRAN SASSO, or
AMANDA as the target. We find that the SOUDAN and GRAN SASSO
possibilities would probe some of the possible $\nu_e \rightarrow
\nu_\tau$ oscillation region. We also compare this model
to another four-neutrino mass matrix parameterization~\cite{Mohap} which
has been proposed to explain the data, and discuss their similarities
and differences.

\underline{\it LSND.}
The LSND experiment~\cite{LSND} searches for $\bar\nu_\mu \rightarrow
\bar\nu_e$ oscillations from $\mu^+$ decay at rest (DAR) and for
$\nu_\mu\rightarrow\nu_e$ oscillations from $\pi^+$ decay in flight
(DIF). The DAR data has higher statistics, but the allowed regions for
the two processes are in good agreement and suggest
$\nu_\mu\rightarrow\nu_e$ vacuum oscillation
parameters that lie along the line segment described by
\begin{equation}
0.3 {\rm~eV}^2 \le \delta m^2_{\rm LSND} =
{0.030 {\rm~eV}^2 \over (A_{\rm LSND})^{0.7}} \le 2.0 {\rm~eV}^2 \;.
\label{lsnddata}
\end{equation}
Larger values for
$\delta m^2_{\rm LSND}$ are excluded at 90\%~C.L. by the BNL
E-776~\cite{E-776} and KARMEN~\cite{KARMEN} oscillation search
experiments, and smaller values for $\delta m^2_{\rm LSND}$ are
excluded by the Bugey reactor experiment, which looks for $\bar\nu_e$
disappearance~\cite{Bugey}. If only 99\%~C.L. exclusion is required,
$\delta m^2_{\rm LSND}$ as high as 10~eV$^2$ is allowed for
$A_{\rm LSND} \simeq .0025$; values of $\delta m^2_{\rm LSND} >
10$~eV$^2$ are excluded by the NOMAD experiment~\cite{NOMAD},
while values above 3 eV$^2$ are disfavored by the r--process mechanism
of heavy element nucleosynthesis in supernovae~\cite{rprocess}.

\underline{\it Atmospheric.}
The atmospheric neutrino experiments measure $\nu_\mu$ and $\nu_e$ (and
their antineutrinos) created when cosmic rays interact with the Earth's
atmosphere. One expects about twice as many muon neutrinos as electron
neutrinos from the resulting cascade of pion and other meson decays.
Several experiments~\cite{atmos,SuperK} obtain a $\nu_\mu/\nu_e$ ratio
that is about 0.6 of the value expected from detailed theoretical
calculations of the flux~\cite{flux}.  The Super-Kamiokande experiment
has collected the most data and a preliminary analysis indicates that
their results can be explained as $\nu_\mu\rightarrow\nu_\tau$
oscillations with~\cite{review,SuperK,oldatmos}
\begin{equation}
3\times10^{-4} {\rm~eV}^2 \le \delta m^2_{\rm atm.}
\le 7\times10^{-3} {\rm~eV}^2 \,, \quad
0.8 \le A_{\rm atm.} \le 1.0 \;,
\label{atmdata}
\end{equation}
with the high end of each range favored. Although
$\nu_\mu\rightarrow\nu_s$ oscillations (where $\nu_s$ is sterile) could
in principle explain the atmospheric data, big-bang nucleosynthesis
excludes this possibility~\cite{BBN} unless the chemical potential of
the neutrinos is modified~\cite{chempot}. Independent of flux normalization
considerations, the $\nu_\mu\rightarrow\nu_e$ oscillation channel is
strongly disfavored by the zenith angle distributions of the
data~\cite{SuperK}. The recent CHOOZ $\bar\nu_e$ disappearance
experiment also excludes $\bar\nu_e \rightarrow \bar\nu_\mu$
oscillations with large mixing $A \gsim 0.2$ for $\delta m^2 \geq 10^{-3}
{\rm eV}^2$~\cite{CHOOZ}.

\underline{\it Solar.}
The solar neutrino experiments~\cite{solar} measure $\nu_e$ created in
the sun. There are three types of experiments, $\nu_e$ capture in Cl in
the Homestake mine, $\nu_e-e$ scattering at Kamiokande and
Super-Kamiokande, and $\nu_e$ capture in Ga at SAGE and GALLEX; each is
sensitive to different ranges of the solar neutrino spectrum and
measures a suppression from the expectations of the standard solar
model (SSM)\cite{SSM}.
Matter-enhanced MSW oscillations of $\nu_e \rightarrow \nu_\mu,
\nu_\tau$~\cite{MSW}, or $\nu_s$~\cite{MSWsterile} are sufficient to
explain the data.  For $\nu_e\rightarrow\nu_s$ the allowed parameter
region~\cite{hata} is bounded by
\begin{equation}
3.5\times10^{-6} {\rm~eV}^2 \leq
\delta m^2_{\rm sol.} \leq
7.5\times10^{-6} {\rm~eV}^2 \,, \quad
2.5\times10^{-3} \leq A_{\rm sol.} \leq 1.6\times10^{-2} \;.
\label{soldata}
\end{equation}
%
%
%
%
The allowed regions for small--angle MSW 
$\nu_e \rightarrow \nu_\mu,\nu_\tau$ solutions are slightly smaller.
However, if the LSND data is to be explained by
$\nu_\mu \rightarrow \nu_e$ oscillations at the relatively large mass
scale indicated in Eq.~\ref{lsnddata}, and the atmospheric data by
$\nu_{\mu} \rightarrow \nu_{\tau}$ oscillations with the much smaller
scale of Eq.~\ref{atmdata}, then the solar neutrino data would appear to
suggest $\nu_e \rightarrow \nu_s$ since the $\delta m^2_{e\tau}$ scale,
which is also given by Eq.~\ref{lsnddata}, is not consistent with
Eq.~\ref{soldata}. It is necessary that the eigenmass
$m_0$ associated predominantly with $\nu_s$ be heavier than
the mass $m_1$ associated predominantly with $\nu_e$ so that it is
$\nu_e$ rather than ${\bar\nu}_e$ that is resonant in the sun, which
requires
\begin{equation}
\delta m^2_{01} = m_0^2 - m_1^2 > 0 \;.
\label{positive}
\end{equation}

\underline{\it Complete description of the data.}
In order to explain all of the above data, one needs a model which
includes the three different two-neutrino solutions described above.
The appropriate mass scales $\delta m^2_{01}$ for MSW solar,
$\delta m^2_{32}$ for atmospheric, and $\delta m^2_{21} \simeq
\delta m^2_{31} \simeq \delta m^2_{31} \simeq \delta m^2_{31}$
for LSND oscillations are provided by the mass hierarchy
$m_1^2 \lsim m_0^2 \ll m_2^2 \simeq m_3^2$. Taking the small--angle
solar MSW solution, the required
oscillation amplitude hierarchy is $A_{es} \simeq A_{e\mu}
\simeq A_{e\tau} \ll A_{\mu\tau} \simeq 1$.

\underline{\it Mass matrix ansatz.}
To describe the above oscillation phenomena, we consider the 
neutrino mass matrix
\begin{equation}
M = m \left( \begin{array}{cccc}
\eps_1^2        & \eps_1^2 \eps_2 & 0        & 0\\
\eps_1^2 \eps_2 & 0               & 0        & \eps_3\\
0               & 0               & \eps_4^2 & 1\\
0               & \eps_3          & 1        & \eps_4^2
\end{array} \right) \;,
\label{m}
\end{equation}
presented in the ($\nu_s, \nu_e, \nu_{\mu}, \nu_{\tau}$) basis. The mass
matrix $M$ contains five parameters ($m$, $\eps_1$, $\eps_2$, $\eps_3$,
$\eps_4$), just enough to incorporate the required three mass differences
and two small oscillation amplitudes $A_{e\mu}$ and $A_{es}$. The large
amplitude $A_{\mu\tau}$ does not require a sixth parameter in our model,
because the structure of the $\nu_{\mu}$--$\nu_{\tau}$ submatrix
naturally gives maximal mixing here (more on this below).  We note that
changing the position of $\eps_3$ from the $M_{e\tau}$ element to the
$M_{e\mu}$ element would cause the $\nu_\tau$ to oscillate into $\nu_e$
instead of into $\nu_\mu$. If nonzero terms are introduced at both the
$M_{e\tau}$ and $M_{e\mu}$ positions, then the physics changes: both
$\nu_{\mu}$ and $\nu_{\tau}$ would mix with $\nu_e$ at the LSND
scale, and the $\nu_e-\nu_s$ mixing angle is also affected\footnote
{The other zero terms could be taken as nonzero without
changing the phenomenology discussed here as long as they are
small compared to $\eps_1^2$.
Inclusion of a small nonzero $M_{ee}$ term merely increases the tiny 
eigenmass $m_1$, while a small nonzero $M_{s\mu}$ or $M_{s\tau}$
gives the sterile neutrino a larger but nevertheless unobservable
mixing with $\nu_\mu$ and $\nu_\tau$.}.
Here we choose to take the minimal $M$ needed to describe the data
and determine the consequences.

For simplicity, we have taken the mass matrix to be real and symmetric;
then $M$ is diagonalized by an orthogonal matrix $U$. Since $U$ is real,
there is no CP violation (which should be small anyway since observable
CP--violation requires more than one large mixing angle, while the data
seems to indicate just one). The $\eps_j$ are assumed to be small and of
the same order of magnitude; phenomenologically they turn out to be
within a factor of two of each other.

The smallness of $M_{es}/|M_{ss}-M_{ee}|=\eps_2$ is designed to yield
the small-angle MSW solution for the
sun. Simple changes in the 2x2 $\nu_s-\nu_e$ sub-block of $M$ would
allow us to also consider the large--angle MSW solar solution, but since
large mixing of sterile with active neutrinos is disfavored by the
solar data~\cite{hata} and big-bang nucleosynthesis~\cite{BBN} for
these $\delta m^2$ values, we do not pursue this option here.

To a good approximation, the eigenvalues of the mass matrix in
Eq.~\ref{m} are
\begin{equation}
m_0 \simeq m\eps_1^2\,, \quad
m_1 \simeq m(\eps_3^2 \eps_4^2 - \eps_1^2 \eps_2^2)\,, \quad 
\ts m_{2,3} = \mp m \left( 1 \mp \eps_4^2 + {1\over2}\eps_3^2\right)\;,
\end{equation}
which shows the desired hierarchy. The small relative mass splitting of
the heavier masses $m_2,m_3$ is governed entirely by the parameter
$\eps_4^2$. Defining $\delta m^2_{ij} = m_i^2 - m_j^2$, the LSND
$\nu_\mu \rightarrow \nu_e$ oscillations are driven by the scale $m^2
\simeq \delta m^2_{21} \simeq \delta m^2_{31} \simeq \delta m^2_{20}
\simeq \delta m^2_{30}$, the atmospheric $\nu_\mu\rightarrow\nu_\tau$
oscillations are determined by $\delta m^2_{32} \simeq 4m^2\eps_4^2$,
and the solar $\nu_e \rightarrow \nu_s$ oscillations are determined by
$\delta m^2_{01} \simeq m^2 \eps_1^4$. The charged-current eigenstates
are approximately related to the mass eigenstates by
\begin{equation}
\left( \begin{array}{c}
\nu_s \\ \nu_e \\ \nu_\mu \\ \nu_\tau
\end{array} \right) = U
\left( \begin{array}{c}
\nu_0 \\ \nu_1 \\ \nu_2 \\ \nu_3
\end{array} \right) =
\left( \begin{array}{cccc}
1                                     & -\eps_2
& -{1\over\sqrt2}\eps_1^2\eps_2\eps_3 & {1\over\sqrt2}\eps_1^2\eps_2\eps_3 \\
\eps_2                                & 1
& {1\over\sqrt2}\eps_3                & {1\over\sqrt2}\eps_3 \\
-\eps_2\eps_3                         & -\eps_3
& {1\over\sqrt2}                      & {1\over\sqrt2} \\
\eps_2\eps_3(\eps_4^2-\eps_1^2)       & \eps_3\eps_4^2
& -{1\over\sqrt2}                     & {1\over\sqrt2}
\end{array} \right)
\left( \begin{array}{c}
\nu_0 \\ \nu_1 \\ \nu_2 \\ \nu_3
\end{array} \right) \;.
\label{U}
\end{equation}
Unitarity holds to first order in the $\eps_j$: $U U^{\dag} = 1 +
{\cal O}(\eps_j^{2})$. Note that $\nu_0$ and $\nu_1$ couple
predominantly to $\nu_s$ and $\nu_e$, respectively, as desired.  
The near--degenerate $\nu_2$ and $\nu_3$ are seen to consist primarily
of nearly equal mixtures of $\nu_\mu$ and $\nu_\tau$.  
These results are shown schematically in Fig.~1.

\underline{\it Oscillation probabilities.}
With real--valued $U$, the vacuum oscillation probabilities are,
in general, given by~\cite{VBreal}
\begin{equation}
P(\nu_\alpha\to \nu_\beta) = \delta_{\alpha\beta}
- 4 \sum_{i<j} U_{\alpha i} U_{\beta i}
U_{\alpha j} U_{\beta j} \sin^2 \Delta_{ji} \;,
\end{equation}
where $\Delta_{ji} \equiv \delta m_{ji}^2 L/4E = 1.27
(\delta m^2/{\rm eV}^2) (L/{\rm km})/(E/{\rm GeV})$. For the mixing in
Eq.~\ref{U}, the off-diagonal vacuum oscillation probabilities, to
leading order in $\eps_j$ for each $\Delta_{ij}$ and ignoring
oscillations smaller than ${\cal O}(\eps_j^4)$, are given by
\begin{eqnarray}
P(\nu_e\to\nu_\mu) &\simeq&
\eps_3^2 \left( 2\sin^2\Delta_{21} + 2\sin^2\Delta_{31}
- \sin^2\Delta_{32} \right) \nonumber\\
&& {}+ \eps_2^2 \eps_3^2 \left( 2\sin^2\Delta_{20} + 2\sin^2\Delta_{30}
- 4\sin^2\Delta_{01} \right) \;,
\label{prob1} \\
P(\nu_e\to\nu_\tau) &\simeq&
\eps_3^2 \sin^2\Delta_{32}  + 2\eps_3^2\eps_4^2 \left(  
\sin^2\Delta_{21} - \sin^2\Delta_{31} \right) \;,
\label{prob2} \\
P(\nu_\mu\to\nu_\tau) &\simeq&
\sin^2\Delta_{32} + 2\eps_3^2\eps_4^2  
\left(\sin^2\Delta_{31} - \sin^2\Delta_{21} \right) \;,
\label{prob3} \\
P(\nu_e\to\nu_s) &\simeq& 4\eps_2^2 \sin^2\Delta_{01} \;,
\label{prob4} \\
P(\nu_\mu\to\nu_s) &\simeq& 4\eps_2^2\eps_3^2 \sin^2\Delta_{01} \;,
\label{prob5}
\end{eqnarray}
where $\Delta_{01} \ll \Delta_{32} \ll \Delta_{20} \simeq \Delta_{30}
\simeq \Delta_{21} \simeq \Delta_{31}$ due to the spectrum of the
neutrino mass eigenvalues.

{\sl For small $L/E$} only the leading oscillations 
$\Delta_{20} \simeq \Delta_{21}  
\simeq \Delta_{30} \simeq \Delta_{31}$ contribute, and the only
appreciable oscillation probability is
\begin{equation}
P(\nu_e\to\nu_\mu) \simeq 4\eps_3^2 \sin^2\Delta \;,
\label{lsndosc}
\end{equation}
where $\Delta\equiv m^2 L/4E$. From Eq.~\ref{lsndosc} we can fix
two model parameters
\begin{equation}
\delta m^2_{\rm LSND} = m^2 \;, \quad A_{\rm LSND} = 4 \eps_3^2 \;.
\label{lsndcon}
\end{equation}
The vacuum oscillation length associated with the LSND $\delta m^2$
scale is
\begin{equation}
\lambda_v = 4\pi E/\delta m^2 =
2.5 {\rm~km} (E/{\rm GeV})(\delta m^2_{\rm LSND}/{\rm eV}^2)^{-1} \;.
\end{equation}

{\sl For $L/E$ typical to atmospheric or long baseline neutrino experiments}, 
the oscillations in $\Delta$ assume their average values. The $\Delta_{32}$  
oscillation is now evident, and the non-negligible oscillation
probabilities in vacuum are
\begin{eqnarray}
P(\nu_e\to\nu_\mu) &\simeq& \eps_3^2 \left(2 - \sin^2\Delta_{32}\right) \;,
\label{lb1} \\
P(\nu_e\to\nu_{\tau}) &\simeq& \eps_3^2 \sin^2\Delta_{32} \;,
\label{lb2} \\
P(\nu_\mu\to\nu_\tau) &\simeq& \sin^2\Delta_{32} \;.
\label{lb3}
\end{eqnarray}
{}From Eq.~\ref{lb3}
\begin{equation}
\delta m^2_{\rm atm.} = \delta m^2_{32} \simeq 4 m^2 \eps_4^2 \;, \quad
A_{\rm atm.} = 1 \;,
\end{equation}
which determines another parameter of the model.
The model automatically gives maximal $\nu_\mu\rightarrow\nu_\tau$
oscillations for atmospheric neutrinos, while oscillations in other
channels are suppressed. The $\nu_\mu$--$\nu_\tau$ maximal mixing
is natural in the sense that it results from the large value of the
$M_{\mu\tau}$ matrix element relative to the diagonal $M_{\mu\mu}$
and $M_{\tau\tau}$ elements, without any need for fine tuning of the
difference $|M_{\mu\mu}-M_{\tau\tau}|$. The oscillation length
resulting from the $\delta m^2_{32}$ scale is 
\begin{equation}
\lambda_v = 500 {\rm~km} (E/{\rm GeV})
(\delta m^2_{\rm atm}/5\times 10^{-3} {\rm eV}^2)^{-1} \;.
\end{equation}

Finally, {\sl for very large} $L/E \gg
(\delta m^2_{\rm atm.}/{\rm eV}^2)^{-1}$~km/GeV, $\sin^2\Delta_{32}$
averages to ${1\over2}$ and the appreciable oscillations in vacuum
are (to leading order in the $\eps_j$)
\begin{eqnarray}
P(\nu_e\to\nu_s) &\simeq& 4 \eps_2^2 \sin^2\Delta_{01} \;,
\label{sol1} \\
P(\nu_e\to\nu_\mu) &\simeq& {3\over2} \eps_3^2 \;,
\label{sol2} \\
P(\nu_e\to\nu_\tau) &\simeq& {1\over2} \eps_3^2 \;,
\label{sol3} \\
P(\nu_\mu\to\nu_\tau) &\simeq& {1\over2} \;,
\label{sol4} \\
P(\nu_\mu\to\nu_s) &\simeq&  4\eps_2^2\eps_3^2 \sin^2\Delta_{01} \;.
\label{sol5}
\end{eqnarray}
The solar data can then be explained with the usual
MSW matter--enhanced mechanism (including the proper sign of
$\delta m^2_{01}$ in Eq.~\ref{positive}) if the parameters in vacuum
satisfy
\begin{equation}
\delta m^2_{\rm sol.} = \delta m^2_{01} \simeq 4 m^2 \eps_1^4 \;, \quad
A_{\rm sol.} = 4 \eps_2^2 \;.
\end{equation}

Summarizing the above analysis, the model parameters are related to the
observables by
\begin{equation}
m^2 = \delta m^2_{\rm LSND} \,, \quad \eps_1^4 = {\delta m^2_{\rm sol.}\over
\delta m^2_{\rm LSND}} \,, \quad \eps_2^2 = {A_{\rm sol.}\over 4} \,, \quad
\eps_3^2 = {A_{\rm LSND}\over4} \,, \quad \eps_4^2 = {\delta m^2_{\rm atm.}  
\over 4\delta m^2_{\rm LSND}} \;.
\end{equation}
%
%
%
%
%
For the specific values $\delta m^2_{\rm LSND}= 2\rm~eV^2$,
$A_{\rm LSND} = 2.5\times10^{-3}$,
$\delta m^2_{\rm atm.} = 5\times10^{-3}\rm~eV^2$, $A_{\rm atm.} = 1$,
$\delta m^2_{\rm sol.} = 4\times10^{-6}\rm~eV^2$,
and $A_{\rm sol.} = 1\times10^{-2}$, we obtain
\begin{equation}
m = 1.4 {\rm~eV} \,, \quad
\eps_1 = 0.038 \,, \quad
\eps_2 = 0.050 \,, \quad
\eps_3 = 0.025 \,, \quad
\eps_4 = 0.025 \,.
\end{equation}
The corresponding neutrino mass eigenvalues are (in eV)
\begin{equation}
m_0 = 2\times 10^{-3} \,, \quad
m_1 = 4\times 10^{-6} \,, \quad
{1\over2}(m_3 + m_2) \simeq 1.4 \,, \quad
{1\over2}(m_3 - m_2) \simeq 9\times10^{-4} \;.
\end{equation}
For these masses $\sum m_\nu \approx 3$~eV, which according to recent
work on early universe formation of the largest structures
provides an ideal hot dark matter component~\cite{hdm}.

If instead we use the lowest allowed mass scale for the LSND experiment
we obtain $\delta m^2_{\rm LSND} = 0.3\rm~eV^2$ and $A_{\rm LSND} =
4\times10^{-2}$, in which case
\begin{equation}
m = 0.55 {\rm~eV}\,, \quad
\eps_1 = 0.060 \,, \quad
\eps_2 = 0.050 \,, \quad
\eps_3 = 0.10 \,, \quad
\eps_4 = 0.065 \,,
\end{equation}
with corresponding mass eigenvalues (in eV)
\begin{equation}
m_0 = 2\times 10^{-3} \,, \quad
m_1 = 2\times 10^{-5} \,, \quad
{1\over2}(m_3 + m_2) \simeq 0.55 \,, \quad
{1\over2}(m_3 - m_2) \simeq 2.3\times10^{-3} \;.
\end{equation}
In either of the above examples, the $\delta m^2$ scale for the
atmospheric neutrino oscillation can be adjusted simply by varying
$\eps_4$.  Also in either case, the two heaviest masses provide relic
neutrino targets for a mechanism that may generate the cosmic ray air
showers observed above $\gsim 10^{20}$~eV~\cite{relic}.

\underline{\it Model predictions.}
The model is constructed to provide the effective two-neutrino
oscillation solutions for the LSND, atmospheric and solar data. The
Solar Neutrino Observatory (SNO)~\cite{SNO}, which can measure both
charge-current (CC) and neutral-current (NC) interactions, will be
able to test the $\nu_e \rightarrow \nu_s$ solar oscillation hypothesis:
in the sterile case the CC/NC ratio in SNO would be unity and both CC
and NC rates would be suppressed from the SSM predictions.

Given the order of magnitude of the $\delta m^2_{ij}$ and $U_{\alpha j}$,
observable new phenomenology occurs for $L/E \gg 1$~km/GeV in the
oscillation channels
\begin{eqnarray}
P(\nu_e \rightarrow \nu_\mu) &\simeq&
{1\over4} A_{\rm LSND} (2-\sin^2\Delta_{\rm atm.}) \;,
\label{emu}\\
P(\nu_e \rightarrow \nu_\tau) &\simeq&
{1\over4} A_{\rm LSND} \sin^2\Delta_{\rm atm.} \;,
\label{etau}
\end{eqnarray}
where $A_{\rm LSND}\sim {\cal O}(1\%)$ is the oscillation {\it
amplitude} which describes the LSND results and $\Delta_{\rm atm.}
= 1.27 \delta m^2_{\rm atm.} L/E \sim
(\delta m^2_{\rm atm.}/5\times10^{-3}{\rm~eV}^2)
(L/157{\rm km})({\rm GeV}/E)$ is the oscillation {\it argument} 
which describes the atmospheric neutrino data. In addition to the
$\nu_\mu \rightarrow \nu_e$ oscillations due to $\Delta$ in
Eq.~\ref{lsndosc}, which reach their oscillation-averaged value
of ${1\over2} A_{\rm LSND}$, the model
predicts new oscillations in the $\nu_e \rightarrow \nu_\mu$ and
$\nu_e \rightarrow \nu_\tau$ channels with common oscillation length
determined by $\Delta_{\rm atm.}$ and amplitude given by ${1\over4}
A_{\rm LSND}$.

How can the oscillation probabilities in Eqs.~~\ref{emu} and \ref{etau}
be tested? A list of experiments currently underway or
being planned to test neutrino oscillation hypotheses is given in
Table~1~\cite{experiments}. In each case the oscillation channel
and the parameters which are expected to be tested are shown. Many of
these experiments will not
provide any constraints on the new phenomenology, although many provide
some check on the existing LSND or atmospheric neutrino results (those
that provide tests are noted in the table). The KARMEN
upgrade, Booster Neutrino Experiment (BooNE at
Fermilab), ORLANDO at Oak Ridge, and MINOS
(Fermilab to SOUDAN) can test the LSND $\nu_\mu \rightarrow
\nu_e$ oscillations, and for the experiments that probe $\delta m^2$
significantly below 1~eV$^2$, may be able to detect the contribution
of the additional oscillation due to the $\sin^2\Delta_{32}$ term in
Eq.~\ref{emu}. MINOS and ICARUS also aim
to detect $\nu_\tau$ and should be able to probe some of the atmospheric
neutrino allowed region for $\nu_\mu \rightarrow \nu_\tau$. NOMAD,
CHORUS, and TOSCA at CERN and COSMOS at Fermilab
will test $\nu_\mu\rightarrow\nu_\tau$ oscillations at most down to
$\delta m^2 \approx 0.1 {\rm~eV}^2$ for maximum amplitude; these do not
probe our model as there are no appreciable $\nu_\mu\rightarrow\nu_\tau$
oscillations in that region. Reactor experiments at Palo Verde
in Arizona and with the BOREXINO detector in Europe will test
$\bar\nu_e$ disappearance involving appreciable mixing angles,
but will not test our model since the largest $\bar\nu_e$ vacuum
oscillations are the $A=.04$ level or less.

Long-baseline experiments with an
intense $\nu_e$ or $\bar\nu_e$ neutrino beam which can detect $\tau$'s,
and hence see the $\nu_e\rightarrow\nu_\tau$ oscillations in
Eq.~\ref{etau}, can provide a definitive test of the
new phenomenology of our model. High intensity muon sources~\cite{Geer} can
provide simultaneous high intensity $\nu_\mu$ and $\bar\nu_e$ (or
$\bar\nu_\mu$ and $\nu_e$ for antimuons) beams with well-determined
fluxes, which could then be aimed at a neutrino detector at a distant
site.  It is expected that $\tau$'s will be detected through their
$\mu$ decay mode and that a charge determination can be made, so that
one can tell if the $\tau$ originated from $\nu_\mu \rightarrow \nu_\tau$
or $\bar\nu_e \rightarrow \bar\nu_\tau$ oscillations. Current
proposals~\cite{Geer} consider SOUDAN ($L=$~732~km) or GRAN SASSO
($L=$~9900~km) as the far site from an intense muon source at Fermilab.
These experiments could also observe
$\nu_e \rightarrow \nu_\mu$ oscillations via detection of ``wrong-sign''
muons. The neutrino energies are in the 10-50~GeV range. Assuming that
low backgrounds can be achieved, the sensitivity to $\delta m^2$ is
roughly proportional to the inverse square root of the detector size
(given the same neutrino energy spectrum at the source); the $\delta
m^2$ sensitivity does not depend on detector distance $L$ because
although the flux in the detector falls off with $L^2$, the oscillation
argument grows with $L^2$ for small $\delta m^2 L/E$. For 20~GeV muons
at Fermilab and a 10~kT detector at either SOUDAN or GRAN SASSO, the
single-event $\delta m^2$ sensitivity for $\nu_e \rightarrow \nu_\tau$
oscillations is about $8\times10^{-5}$~eV$^2$ for maximal
mixing~\cite{Geer}. For large $\delta m^2$, the oscillation amplitude
single-event sensitivity is roughly inversely proportional to the
neutrino flux at the detector divided by the detector size; about
$6\times10^{-5}$ for SOUDAN and $10^{-2}$ for GRAN SASSO~\cite{Geer}.
In general, the closer detector has comparable $\delta m^2$ sensitivity
but better $A$ sensitivity.

Our model predicts $\nu_e\rightarrow\nu_\tau$ oscillations with
amplitude $A_{\rm LSND}/4$ (which ranges from 0.0025 to 0.04) and
mass-squared difference of $\delta m^2_{\rm atm.}$ (which ranges from
$3\times10^{-4}$ to $7\times10^{-3}$ eV$^2$). The region of possible
$\nu_e\rightarrow\nu_\tau$ oscillations in our model and the regions
which can be tested at the SOUDAN and GRAN SASSO sites are shown
schematically in Fig.~2, along with the favored parameters for the LSND,
atmospheric neutrino, and solar neutrino oscillations. Such experiments
would be sensitive to some of the $\nu_e \rightarrow
\nu_\tau$ region, though they may not cover the low-mass,
small-amplitude part. These searches would also be able to test the
$\nu_e \rightarrow \nu_\mu$ oscillations in Eq.~\ref{emu} and the
atmospheric $\nu_\mu \rightarrow \nu_\tau$ oscillations. Additionally,
long baseline experiments to the AMANDA~\cite{AMANDA} detector from Fermilab or KEK may be useful in probing oscillations with small $\delta m^2$.

\underline{\it Neutrinoless double--$\beta$ decay.}
{}From the form of U and the mass eigenvalues one can readily see
that neutrinoless double--$\beta$ decay is unobservable in our model.
If neutrinos are Majorana particles, then the $(\beta\beta)_{0\nu}$
decay rate is proportional to 
\begin{equation}
<m_\nu>\equiv|\sum_j U^2_{ej} m_j| \sim m\eps_3^2 < 10^{-2} {\rm eV} \;,
\end{equation}
which is well below the present limit of $\sim$ 0.5 eV~\cite{KK},
and less than improved bounds realizable in the future.  Note that possible
CP--violating relative Majorana phases which we have ignored in our model 
can give smaller $<m_\nu>$ via a cancellation in the leading terms, but
cannot give larger $<m_\nu>$.

\underline{\it Hot dark matter.}
The contribution of the neutrinos to the mass density of the universe is
given by $\Omega_\nu = \sum m_\nu/(h^2 93$~eV), where $h$ is the Hubble
expansion parameter in units of 100 km/s/Mpc~\cite{expansion}; with
$h=0.65$ our model implies $\Omega_\nu \approx 0.05$. An
interesting test of neutrino masses is the Sloan Digital
Sky Survey (SDSS)~\cite{SDSS}. For two nearly degenerate massive
neutrino species, sensitivity down to about 0.2 to 0.9~eV (depending on
$\Omega$ and $h$) is expected, providing coverage of all or part of the
LSND allowed range ($m=$0.55 to 1.4~eV in our model).

\underline{\it Resonant enhancement in matter.}
The curves in Fig.~2 assume vacuum oscillations. In general, large
corrections to oscillations involving $\nu_e$ and $\nu_s$ are possible
due to matter; the $\nu_e$ diagonal element in the effective
mass-squared matrix receives an additional term $2\sqrt2 G_F N_e E$
from the CC interaction, and the $\nu_s$ diagonal element receives the
contribution $\sqrt2 G_F N_n E$ (relative to the active neutrinos)
because it does not have NC interactions, where $N_e$ and $N_n$ are
the electron and neutron number density, respectively. In our model,
however, these corrections do not significantly affect the large
$m_2^2$ and $m_3^2$ mass eigenvalues as long as $E \ll$~1~TeV, and
hence only modify the $\delta m^2_{01}$ oscillation argument. For
$E \ge 0.1$~GeV, the only significant change in the mixing parameters
is that the $\nu_e-\nu_s$ mixing $\eps_2$ in Eqs.~\ref{prob1} to
\ref{prob5} is suppressed. The result is that for small and intermediate
$L/E$ ({\it i.e.}, all experiments described by Eqs.~\ref{lsndosc},
\ref{lb1}, \ref{lb2}, and \ref{lb3}), matter does not appreciably
change the observable phenomenology of the model. For large
$L/E$, such as when $E \lsim 10$~MeV for solar neutrinos, there can
be  maximal $\nu_e-\nu_s$ mixing which changes Eqs.~\ref{sol1},
\ref{sol2}, and \ref{sol5} to
\begin{eqnarray}
P(\nu_e\to\nu_s) &\simeq& \sin^2\Delta_{01} \;,
\label{sol6} \\
P(\nu_e\to\nu_\mu) &\simeq&
\eps_3^2 \left( {3\over2} - \sin^2\Delta_{01} \right) \;,
\label{sol7} \\
P(\nu_\mu\to\nu_s) &\simeq& \eps_3^2 \sin^2\Delta_{01} \;.
\label{sol8}
\end{eqnarray}
In this case the only significant effect of matter (other than the MSW
enhancement of $\nu_e \rightarrow \nu_s$ that leads to the solar
neutrino suppression) is to enhance the $\nu_\mu \rightarrow \nu_s$
oscillations and introduce a new oscillation in the $\nu_e \rightarrow
\nu_\mu$ channel, although the amplitude of these new oscillations
never gets above about 10$^{-2}$. Hence we conclude that the matter
corrections for the mass matrix in Eq.~\ref{m} probably have no
observable consequences.

\underline{\it Other models.}
Are other viable neutrino mixing schemes possible? A
different form for the neutrino mass matrix is
\begin{equation}
M = m \left( \begin{array}{cccc}
\eps_1^2        & \eps_1^2 \eps_2 & 0        & 0 \\
\eps_1^2 \eps_2 & 0               & \eps_3   & 0 \\
0               & \eps_3          & 1        & \eps_4^2 \\
0               & 0               & \eps_4^2 & 1+\eps_5^2
\end{array} \right) \;.
\label{m2}
\end{equation}
This alternate form contains one more parameter, $\eps^2_5$, than the
mass matrix in Eq.~\ref{m}. Fine--tuning of this additional
parameter is necessary to achieve maximal mixing in the $\nu_2-\nu_3$
sector. Eq.~\ref{m2} is a generalization of the particular form used in
Ref.~\cite{Mohap} which has $\eps_5^2 =2\eps_4^2$. Again, zero elements
can be taken nonzero as long as they are very small.
The eigenvalues are given approximately by
\begin{equation}
m_0 \simeq  m\eps_1^2 \,, \quad
m_1 \simeq -m\eps_3^2 \,, \quad
m_{2,3} \simeq m\left[1+{1\over2}(\eps_3^2+\eps_5^2) \mp
\sqrt{{1\over4}(\eps_3^2-\eps_5^2)^2+\eps_4^4}\right] \;.
\end{equation}
The charged current eigenstates are approximately related to the
mass eigenstates by
\begin{equation}
\left( \begin{array}{c}
\nu_s \\ \nu_e \\ \nu_\mu \\ \nu_\tau
\end{array} \right) = U
\left( \begin{array}{c}
\nu_0 \\ \nu_1 \\ \nu_2 \\ \nu_3
\end{array} \right) =
\left( \begin{array}{cccc}
1                         & -\beta\eps_2
& \eps_1^2\eps_2\eps_3 c  & \eps_1^2\eps_2\eps_3 s \\
\beta\eps_2               & 1
& \eps_3 c                & \eps_3 s \\
-\beta\eps_2\eps_3        & -\eps_3
& c                       & s \\
\beta\eps_2\eps_3\eps_4^2 & \eps_3\eps_4^2
& -s                      & c
\end{array} \right)
\left( \begin{array}{c}
\nu_0 \\ \nu_1 \\ \nu_2 \\ \nu_3
\end{array} \right) \;.
\end{equation}
where
\begin{equation}
\beta\equiv\eps_1^2/(\eps_1^2+\eps_3^2) \;,
\end{equation}
and
\begin{equation}
s \equiv \sin\theta_{\mu\tau} = {1\over\sqrt{2}}
\left[ 1 + {{1\over2}(\eps_3^2-\eps_5^2)\over
\sqrt{ {1\over4}(\eps_3^2-\eps_5^2)^2+\eps_4^4}} \right]^{1/2}
\end{equation}
determines the $\nu_\mu-\nu_\tau$ mixing and $c \equiv
\cos\theta_{\mu\tau}$. We also have
\begin{equation}
\delta m^2_{sol.}=m^2_0 -m^2_1 \simeq m^2(\eps_1^4-\eps_3^4) \,, \quad
A_{sol.} \simeq 4\beta^2\eps_2^2 \;.
\end{equation}
Note that $m_1$ cannot be larger than $m_0$, since this would lead to a
negative $\delta m^2_{\rm sol.}$ and would imply that $\bar\nu_e$ and
not $\nu_e$ undergoes an MSW enhancement, contrary to the data.  Hence
$\eps_1$ must be larger than $\eps_3$ with this matrix form, which in
turn limits $\beta$ to the interval $[0.5,1]$.  Then $\eps_2$ must be
raised by the factor $\beta^{-2}$ relative to the matrix form in
Eq.~\ref{m2} to give the proper MSW mixing. The LSND constraints are
the same as in Eq.~\ref{lsndcon}.  In the $\nu_\mu-\nu_\tau$
sector, the parameters are determined by
\begin{equation}
\delta m^2_{\rm atm.} \simeq
2 m^2 \sqrt{(\eps_3^2 - \eps_5^2)^2 + 4\eps_4^4} \,, \quad
A_{\rm atm.} \simeq {4\eps_4^4\over
(\eps_3^2 - \eps_5^2)^2 + 4\eps_4^4} \;.
\end{equation}
In this scenario the $\eps_3$ term significantly affects
not only the lightest mass eigenvalue but also the mass splitting and
mixing angle of the two heavy states.  Consequently, some fine--tuning
is necessary to achieve the proper phenomenology.  Maximal mixing occurs
only when $\eps_3\simeq\eps_5$, in which case $\delta m^2_{\rm atm.} = 4
m^2 \eps_4^2$ just like our previous scenario. However, in the
absence of such fine--tuning, submaximal mixing is probable, in which
case a different value for $\eps_4$ is required to generate the correct
$\delta m^2_{\rm atm.}$.

We may again solve for the parameters
directly in terms of the observables.  In the present case there are
six parameters and six observables including the (now) possibly
non--maximal amplitude for atmospheric oscillations.
The result is 
\begin{equation}
m^2=\delta m^2_{\rm LSND}\,, \quad
\eps^4_1=\frac{\delta m^2_{\rm sol.}}{\delta m^2_{\rm LSND}}
+\frac{A^2_{\rm LSND}}{16},
\end{equation}
\begin{equation}
\eps_2^2=\frac{A_{\rm sol.}}{4}
\left[1+\frac{A_{\rm LSND}\sqrt{\delta m^2_{\rm LSND}}}
{\sqrt{A^2_{\rm LSND}\delta m^2_{\rm LSND}+16\delta m^2_{\rm sol.}}}\right]\,,
\quad
\eps_3^2=A_{\rm LSND}/4\,,
\end{equation}
\begin{equation}
\eps_4^2=\frac{\sqrt{A_{\rm atm.}}\delta m^2_{\rm atm.}}
{4\delta m^2_{\rm LSND}}\,, \quad
\eps_5^2-\eps_3^2=\pm \frac{1}{2}\sqrt{1-A_{\rm atm.}}\,
\frac{\delta m^2_{\rm atm.}}{\delta m^2_{\rm LSND}}\,.
\end{equation}
Since the matrix form in Eq.~\ref{m2} requires some fine--tuning to
explain the data, some higher order terms must be retained in the
expressions for the parameters.

Using the same input parameters as before, 
including maximal mixing in the atmospheric neutrino experiments
(which implies $\eps_5=\eps_3$), 
we find for the largest $\delta m^2_{\rm LSND}$ solution
\begin{equation}
m = 1.4 {\rm~eV} \,, \quad
\eps_1 = 0.039 \,, \quad
\eps_2 = 0.070 \,, \quad
\eps_3 = \eps_5 = 0.025 \,, \quad
\eps_4 = 0.025 \,,
\label{param3}
\end{equation}
with mass eigenvalues (in eV)
\begin{equation}
m_0 = 2.1\times10^{-3} \,, \quad
m_1 = 0.9\times10^{-3} \,, \quad
{1\over2}(m_3 + m_2) \simeq 1.4 \,, \quad
{1\over2}(m_3 - m_2) \simeq 9\times10^{-4} \;.
\label{mass3}
\end{equation}
For the smallest $\delta m^2_{\rm LSND}$ solution, we obtain
\begin{equation}
m = 0.55 {\rm~eV} \,, \quad
\eps_1 = 0.103 \,, \quad
\eps_2 = 0.070 \,, \quad
\eps_3 = \eps_5 = 0.100 \,, \quad
\eps_4 = 0.065 \,,
\label{param4}
\end{equation}
with masses (in eV)
\begin{equation}
m_0 = 5.8\times10^{-3} \,, \quad
m_1 = 5.5\times10^{-3} \,, \quad
{1\over2}(m_3 + m_2) \simeq 0.55 \,, \quad
{1\over2}(m_3 - m_2) \simeq 2.3\times10^{-3} \;.
\label{mass4}
\end{equation}
In Eq.~\ref{param4} some fine--tuning between $\eps_5$ and $\eps_3$
(to the 3\% level) is needed for $\delta m^2_{\rm sol.}$ to have the
correct sign and magnitude. In either Eq.~\ref{param3} or \ref{param4}
the mass scale for the atmospheric neutrino oscillation can also be
adjusted simply by varying $\eps_4$ (for maximal mixing), or by
adjusting $\eps_4$ and $\eps_5$ (for non-maximal mixing). The only new
phenomenology is again $\nu_e\rightarrow\nu_\mu,\nu_\tau$ for $L/E \gg
1$~km/GeV, except that $\sin^2\Delta_{\rm atm.}$ in Eqs.~\ref{emu} and
\ref{etau} is now replaced by $A_{\rm atm.}\sin^2\Delta_{\rm atm.}$
when $A_{\rm atm.} \ne 1$. The possible $\nu_e\rightarrow\nu_\tau$
oscillation amplitude is reduced by a factor $A_{\rm atm.}$ (which is
apparently 0.8 or higher), which shifts the predicted region in Fig.~2
slightly to the left; otherwise this model is very similar to the model
of Eq.~\ref{m}.

\underline{\it Summary.}
In this letter we have presented a four-neutrino
model with three active neutrinos and one sterile neutrino which
naturally has maximal $\nu_\mu\rightarrow\nu_\tau$ oscillations of
atmospheric neutrinos and can also explain the solar neutrino and LSND
results. The model predicts $\nu_e \rightarrow \nu_\tau$ and
$\nu_e \rightarrow \nu_\mu$ oscillations in long--baseline
experiments with $L/E \gg 1$~km/GeV with amplitudes that are determined
by the LSND oscillation amplitude and $\delta m^2$ scale determined by
the oscillation scale of atmospheric neutrinos.  Neutrino
beams from an intense muon source at Fermilab or KEK with a detector at the
SOUDAN or GRAN SASSO sites may be able to test part of the parameter
region for these oscillations channels.

\underline{\it Acknowledgements.}
We thank K. Hagiwara and R.J.N. Phillips for discussions. This work was
supported in part by the U.S. Department of Energy, Division of High
Energy Physics, under Grants No.~DE-FG02-94ER40817 and No.~DE-FG02-95ER40896 and in part by the University of Wisconsin Research Committee with funds granted by the Wisconsin Alumni Research Foundation.

\newpage

\begin{table}
\caption{Current and planned neutrino oscillation experiments. Check-marks denote accessible oscillation channels. The $\delta m^2$ and $\sin^22\theta$ sensitivies are given.}
\def\ds{\displaystyle}
\vbox{\footnotesize
\tabcolsep=.5em
\begin{tabular}{lcccccccccc}
&&&&&&&&& \multicolumn{2}{c}{Test Model?}\\
Experiment&
$\ds\nu_\mu\atop{\ds\downarrow\atop\ds\nu_e}$& $\ds\nu_\mu\atop{\ds\downarrow\atop\ds\nu_\tau}$&
$\ds\nu_e\atop{\ds\downarrow\atop\ds\nu_\tau}$&
$\ds\nu_e\atop{\ds\downarrow\atop\ds\nu_e}$&
$\ds\delta m^2\atop\bigskip\rm (eV^2)$& $\sin^22\theta$&
$\ds\rm Test\atop\ds\rm LSND?$&$\ds\rm Test\atop\ds\rm Atmos?$&
$\ds\nu_\mu\atop{\ds\downarrow\atop\ds\nu_e}$&
$\ds\nu_e\atop{\ds\downarrow\atop\ds\nu_\tau}$
\\[3mm]
\hline
BOONE&  $\surd$& & & & $10^{-2}$& $6\times10^{-4}$&
$\surd$&  & &   \\
BOREXINO& & & &  $\surd$& $10^{-6}$& 0.4& & & &  \\
CHORUS& & $\surd$& & & 0.3& $2\times10^{-4}$& & & & \\
COSMOS& & $\surd$& & & 0.1& $10^{-5}$& &  & &   \\
ICARUS, NOE,& $\surd$& $\surd$& & & $3\times10^{-3}$& $4\times10^{-2}$&  & p&  \\[-2mm]
AQUA-RICH, OPERA\\
KARMEN& $\surd$& & & & $4\times10^{-2}$& $10^{-3}$& $\surd$& &  &  \\
KamLAND&  $\surd$& & & & $2\times10^{-3}$& 0.2&  & &   & \\
K2K& $\surd$& & & & $2\times10^{-3}$& $5\times10^{-2}$\\
Fermilab/Gran Sasso& $\surd$& $\surd$& $\surd$& & $8\times10^{-5}$& $10^{-2}$& p& $\surd$& p& p \\
Fermilab/Soudan& $\surd$& $\surd$& $\surd$& & $8\times10^{-5}$& $6\times10^{-5}$& $\surd$& $\surd$& $\surd$& p \\
MINOS& $\surd$& $\surd$& & & $10^{-3}$& $10^{-2}$& p& p& p&   \\
NOMAD& & $\surd$& & & 0.5& $3\times10^{-4}$& &  & &   \\
ORLANDO, ESS& $\surd$& & & & $3\times10^{-3}$& $10^{-4}$& $\surd$& &  &  \\
Palo Verde & & & & $\surd$& $10^{-3}$& 0.2&  & & & \\
TOSCA& & $\surd$& & & 0.1& $10^{-5}$&\\
\hline
\end{tabular}
\bigskip

 p = partially
}
\end{table}

\clearpage

\begin{figure}
\centering\leavevmode
\epsffile{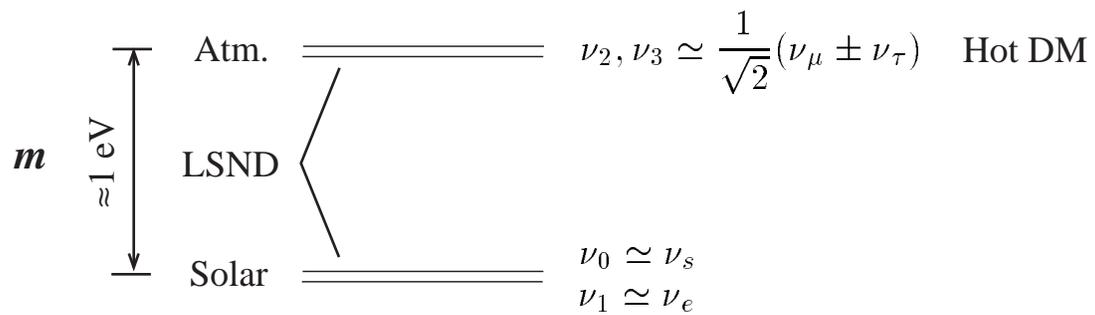}

\caption{Neutrino mass spectrum,
showing the approximate flavor content of each mass eigenstate, and
showing which mass splittings are responsible for the LSND,
atmospheric, and solar oscillations.} 
\end{figure}

\begin{figure}
\centering\leavevmode
\epsfxsize=4.5in\epsffile{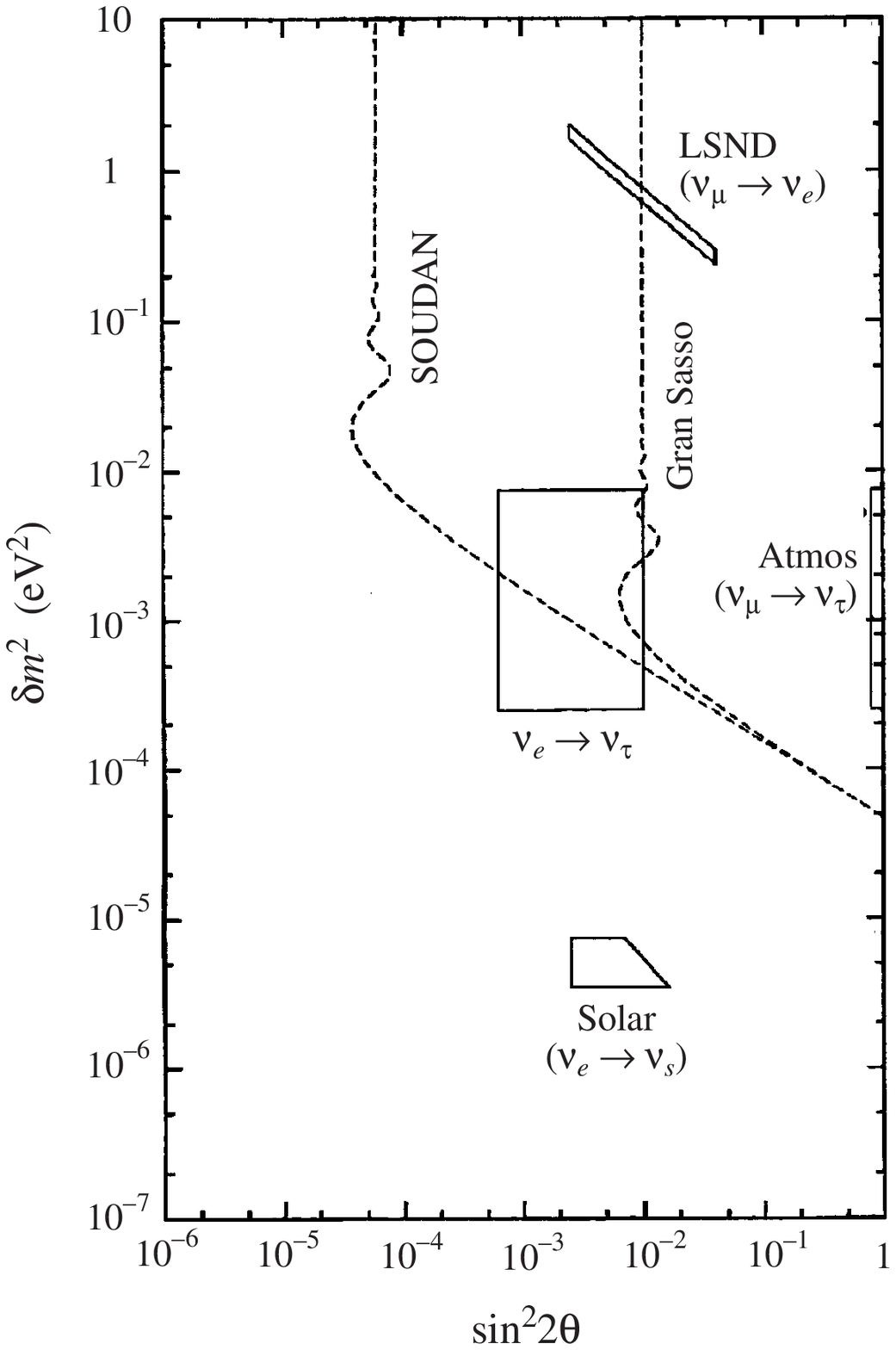}

\caption[]{Predicted region in the effective $\delta
m^2$-$\sin^22\theta$ parameter space for $\nu_e \rightarrow \nu_\tau$
oscillations in the four-neutrino model (solid rectangle), which is
determined by ${1\over4}$ of the
LSND $\nu_\mu \rightarrow \nu_e$ oscillation amplitude and the
atmospheric neutrino $\nu_\mu \rightarrow \nu_\tau$ oscillation $\delta
m^2$ scale. The dashed curves show the potential limits that can be set
by neutrino beams from an intense muon source at Fermilab~\cite{Geer} to
detectors at the SOUDAN and GRAN SASSO sites for muons with energy of
20~GeV. Also shown are the parameters for the solar $\nu_e \rightarrow
\nu_s$ oscillation.}
\end{figure}


\newpage

\begin{thebibliography}{99}
%
\bibitem{review}
For a recent review, see talks at the ITP Conference on Solar Neutrinos:
News About SNUS, Santa Barbara, December 1997 at\\
http://doug-pc.itp.ucsb.edu/online/snu/schedule.html
%
\bibitem{SSM}
J.N. Bahcall and M.H. Pinsonneault, Rev. Mod. Phys. {\bf 67}, 781 (1995).
%
\bibitem{solar}
B.T. Cleveland {\it et al.}, Nucl. Phys. B (Proc. Suppl.) {\bf 38}, 47
(1995);
Kamiokande collaboration, Y. Fukuda {\it et al.}, Phys. Rev. Lett,
{\bf 77}, 1683 (1996);
GALLEX Collaboration, W. Hampel {\it et al.}, Phys. Lett. {\bf B388},
384 (1996);
SAGE collaboration, J.N. Abdurashitov {\it et al.}, Phys. Rev. Lett.
{\bf 77}, 4708 (1996).
%
\bibitem{atmos}
Kamiokande collaboration, K.S. Hirata {\it et al.}, Phys. Lett.
{\bf B280}, 146 (1992); Y. Fukuda {\it et al.}, Phys. Lett.
{\bf B335}, 237 (1994);
IMB collaboration, R. Becker-Szendy {\it et al.}, Nucl. Phys. Proc.
Suppl. {\bf 38B}, 331 (1995);
Soudan-2 collaboration, W.W.M. Allison {\it et al.}, Phys. Lett. {\bf
B391}, 491 (1997).
%
\bibitem{SuperK}
See talk by E. Kearns in Ref.~\cite{review}.
%
\bibitem{oldatmos}
J.G. Learned, S. Pakvasa, and T.J. Weiler, Phys. Lett. {\bf B207}, 79 (1988);
V. Barger and K. Whisnant, Phys. Lett. {\bf B209}, 365 (1988);
M.C. Gonzalez-Garcia, H. Nunokawa, O. Peres, T. Stanev, and J.W.F.
Valle, hep-ph/9712238.
%
\bibitem{LSND}
Liquid Scintillator Neutrino Detector (LSND) collaboration,
C. Athanassopoulos {\it et al.}, Phys. Rev. Lett. {\bf 75}, 2650 (1995);
{\it ibid.} {\bf 77}, 3082 (1996); nucl-ex/9706006.
%
\bibitem{3nu}
G.L. Fogli, E. Lisi, D. Montanino, and G. Scioscia, Phys. Rev. {\bf D
56}, 4365 (1997);
C.Y. Cardall and G.M. Fuller, Nucl. Phys. Proc. Suppl. {\bf 51B}, 259 (1996);
A. Acker and S. Pakvasa, Phys. Lett. {\bf B397}, 209 (1997);
Ernest Ma and Probir Roy, hep-ph/9706309.
%
\bibitem{flat}
P.I. Krastev and S.T. Petcov, Phys. Lett. {\bf B395}, 69 (1997).
%
\bibitem{CHOOZ}
CHOOZ collaboration, M. Apollonio {\it et al.}, hep-ex/9711002.
%
\bibitem{VB80}
V. Barger, P. Langacker, J. Leveille, and S. Pakvasa, Phys. Rev. Lett.
{\bf 45}, 692 (1980);
J.R. Espinosa, hep-ph/9707541; G. Cleaver, M. Cvetic, J.R. Espinosa,
L. Everett, and P. Langacker, hep-ph/9705391.
%
\bibitem{Znunubar}
LEP Electroweak Working Group and SLD Heavy Flavor Group,
D. Abbaneo {\it et al.}, CERN-PPE-96-183, December 1996.
%
\bibitem{models}
D.O. Caldwell and R.N. Mohapatra, Phys. Rev. {\bf D 48}, 3259 (1993);
R. Foot and R.R. Volkas, Phys. Rev. {\bf D 52}, 6595 (1995);
S.M. Bilenky, C. Giunti, and W. Grimus, hep-ph/9711416.
%
\bibitem{Mohap}
R.N. Mohapatra, hep-ph/9711444.
%
\bibitem{BBN}
R. Barbieri and A. Dolgov, Phys. Lett. {\bf B237}, 440 (1990);
K. Enqvist, K. Kainulainen, and M. Thomson, Nucl. Phys. {\bf B373}, 498 (1992);
X. Shi, D.N. Schramm, and B.D. Fields, Phys. Rev. {\bf D 48}, 2563 (1993);
C.Y. Cardall and G.M. Fuller, Phys. Rev. {\bf D 54}, 1260 (1996);
D.P. Kirilova and M.V. Chizhov, hep-ph/9707282.
%
\bibitem{Geer}
S. Geer, hep-ph/9712290.
%
\bibitem{E-776}
L. Borodovsky {\it et al.}, Phys. Rev. Lett. {\bf 68}, 274 (1992).
%
\bibitem{KARMEN}
KARMEN collaboration, B. Bodmann {\it et al.}, Nucl. Phys. {\bf A553},
831c (1993); talk by K. Eitel at 32nd Rencontres de Moriond:
Electroweak Interactions and Unified Theories, Les Arcs, France,
March 1997, hep-ex/9706023.
%
\bibitem{Bugey}
Y. Declais {\it et al.}, Nucl. Phys. {\bf B434}, 503 (1995).
%
\bibitem{NOMAD}
K. Zuber, talk at COSMO'97, Ambleside, England, September 1997, hep-ph/9712378.
%
\bibitem{rprocess}
Y.--Z. Qian {\it et al.}, Phys. Rev. Lett. {\bf 71}, 1965 (1993).
%
\bibitem{flux}
G. Barr, T.K. Gaisser, and T. Stanev, Phys. Rev. {\bf D 39},
3532 (1989);
M. Honda, T. Kajita, K. Kasahara, and S. Midorikawa, Phys. Rev.
{\bf D52}, 4985 (1995);
V. Agrawal, T.K. Gaisser, P. Lipari, and T. Stanev, Phys. Rev.
{\bf D 53}, 1314 (1996);
T.K. Gaisser {\it et al.}, Phys. Rev. {\bf D 54}, 5578 (1996).
%
\bibitem{chempot}
R. Foot and R.R. Volkas, Phys. Rev. Lett. {\bf 75}, 4350 (1995).
%
\bibitem{MSW}
L. Wolfenstein, Phys. Rev. {\bf D 17}, 2369 (1978);
S.P. Mikheyev and A. Smirnov, Yad. Fiz. {\bf 42}, 1441 (1985);
Nuovo Cim. {\bf 9C}, 17 (1986).
%
\bibitem{MSWsterile}
V. Barger, N. Deshpande, P.B. Pal, R.J.N. Phillips, and K. Whisnant,
Phys. Rev. {\bf D 43}, 1759 (1991);
S. Bludman, D.C. Kennedy, and P. Langacker, Nucl. Phys. {\bf B374},
373 (1992).
%
\bibitem{hata}
N. Hata and P. Langacker, hep-ph/9705339.
%
\bibitem{KK}
H.V. Klapdor--Kleingrothaus, hep-ph/9712381.
%
\bibitem{VBreal}
V. Barger, K. Whisnant, D. Cline, and R.J.N. Phillips, Phys. Lett.
{\bf B93}, 194 (1980).
%
\bibitem{hdm}
For a recent discussion see J. Primack, astro-ph/9707285.
%
\bibitem{relic}
T.J. Weiler, hep-ph/9710431.
%
\bibitem{SNO}
E. Norman {\it et al.}, Solar Neutrino Observatory (SNO) collaboration,
in proc. of {\it The Fermilab Conference: DPF 92}, November 1992,
Batavia, IL, ed. by C. H. Albright, P.H. Kasper, R. Raja,
and J. Yoh (World Scientific, Singapore, 1993), p. 1450. 
%
\bibitem{experiments}
For World Wide Web links to more information on these and other
neutrino oscillation experiments, see the Neutrino Oscillation
Industry web page at
http://www.hep.anl.gov/NDK/Hypertext/nuindustry.html.
%
\bibitem{AMANDA}
S. Barwick {\it et al.}, AMANDA collaboration, in proc. XXVIth
International Conference on High Energy Physics, Dallas TX, August 1992,
ed. by James R. Sanford (AIP, New York, 1993), p. 1250; F. Halzen,
astro-ph/9707289.
%
\bibitem{expansion}
E.W. Kolb and M.S. Turner, {\it The Early Universe} (Addison-Wesley,
Reading, 1990). 
%
\bibitem{SDSS}
W. Hu, D.J. Eisenstein, and M. Tegmark, astro-ph/9712057.
%
\end{thebibliography}
\end{document}